\documentclass[11pt,twoside]{article}
\usepackage{./asp2014}

\aspSuppressVolSlug
\resetcounters

\begin{document}

\title{Where Are The Circumbinary Planets of Contact Binaries?}
\author{Osman Demircan,$^1$ and \.{I}brahim Bulut$^1$
\affil{$^1$Department of Space Sciences and Technologies, Faculty of Arts and Sciences, \c{C}anakkale Onsekiz Mart University, \c{C}anakkale, Turkey; \email{demircan@comu.edu.tr}}}

\paperauthor{Osman Demircan}{demircan@comu.edu.tr}{ORCID_Or_Blank}{Author1 Institution}{Department of Space Sciences and Technologies, Faculty of Arts and Sciences, \c{C}anakkale Onsekiz Mart University}{\c{C}anakkale}{State/Province}{TR-17020}{Turkey}
\paperauthor{\.{I}brahim Bulut}{ibulut@comu.edu.tr}{ORCID_Or_Blank}{Author2 Institution}{Astrophysics Research Centre and Observatory, \c{C}anakkale Onsekiz Mart University}{\c{C}anakkale}{State/Province}{TR-17020}{Turkey}

\begin{abstract}

Up to present date, no circumbinary planet around contact binaries were discovered neither by transit method nor by the minima times variation, although they are known having third component stars around. We thus ask: where are the circumbinary planets of contact binaries?

By considering the physical and geometrical parameters we simulated the light curves of contact binaries with possible transiting circumbinary jovian planets.
 
It seems either the circumbinary jovian planets are not formed around contact binaries, probably due to dynamical effects of the binary and third component stars, or they are present but the discovery of such planets were not possible so far due to larger distortions then expected in the photometric data and in the minima times.

\end{abstract}

\section{Introduction}

The disks around stars provide the reservoirs of material out of which planets are formed.  The disks are formed  by lost or accreted material around single stars, but they form at different stages of the evolution if the star is a member of a close binary system: (a) during formation of binary, (b) during Roche Lobe overflow in semi detached binaries, (c) during high rate of mass loss stage, (d) during mass loss from the second Lagrangian point in overcontact binaries, (e) during merging process of contact binaries, (f) during post common envelope stage, and (g) during evaporation of the secondary in X ray binaries. It is thus expected first, second, third, ... generation planets around close binary systems, or around the components of wider binary systems.

Planets in S-type orbits around a component of 44 binaries were found. For these systems the semi-major axis a of the orbit is generally greater than 20 AU \citep[see e.g.][]{d01}.

Circumbinary planets  around 11 post common envelope binaries were also discovered by ground based eclipse timing (ET) observations \citep[see e.g.][]{d02}.

Circumbinary planets in P-type orbits around relatively short period 7 unevolved eclipsing binaries (0.08 < $a$ < 0.23 AU)  were discovered by Kepler's transit observations \citep[see e.g.][]{d03}. All these circumbinary planets (except two outer planets of Kepler 47) were found reside close to the dynamical stability limit boundary (in average \textit{P}$_{p}$ / \textit{P}$_{b}$ = 6.88) which is very close to inner edge of the circumbinary disks where \textit{P}$_{p}$ / \textit{P}$_{b}$ = 4.5.  Due to the difficulty of forming planets in such close orbits, it is believed that they have formed further out in the disk and migrated to their present locations \citep[see e.g.][]{d04}. 

Altough the mass loss may form disks around contact or overcontact binaries, there is no claim of any circumbinary planet around these systems. Either there are no circumbinary planets around contact binaries, or  the observations are not sufficient to detect any such planets, if they exist.

In the present contribution we first simulate the light curves of contact binaries with possible transiting circumbinary jovian planets by considering the physical and geometrical parameters of the circumbinary planets. We than reconsider the ground based ET observations to see if they are capable to detect LITE effects of the circumbinary planets around contact binaries.

\section{Difficulty of Planetary Transits Detection of Contact Binaries}

If exist, the circumbinary planets around contact binaries should have similar properties with those of either unevolved eclipsing binaries (UEB) or those of post common envelope binaries (PCEB). We thus expect orbital periods \textit{P}$_{p}$ of the circumbinary planets around contact binaries would be around 6.88 $\times$ \textit{P}$_{b}$, i.e. not more than a few days in the case of UEB, or not less then ten years in the case of PCEB.  For the simulation we considered a jovian circumbinary planet in one year orbit around the contact binary W UMa system itself. Transit geometry is shown in Fig.1, where the semi-major axis \textit{a}$_{b}$ and \textit{a}$_{p}$ of the binary and planet becomes 2.45 and 260.21 R$_{\odot}$.

The duration of a planetary transit in front of a single star was given by  \citet {d05} which was used by \citet {d10} to derive the following formula for the crossing time \textit{T}$_{GTV}$ of a circumbinary planet in front of a binary system:

\begin{equation}
T_{GTV}=\frac{P_{p}}{\pi}\arcsin (\frac{R_{metastar}}{a_{p}})\frac{\sqrt{1-e_{p}^2}}{1+e_{p}\sin (\omega_{p})},
\end{equation}

where \citet {d10} introduced the quantity \textit{R}$_{metastar}$ representing the maximum extent of the binary's orbit projected on to the sky. The other symbols in equation (1) have their usual meaning for planetary orbit.

For contact binaries, it can be estimated that (\textit{R}$_{metastar}$ $\simeq$ \textit{a}$_{b}$ + \textit{R}$_{1}$ + \textit{R}$_{2}$) where the summation of the mean radii (\textit{R}$_{1}$ + \textit{R}$_{2}$) of the component stars for marginal contact binaries is about 0.75\textit{a}$_{b}$ which may increase up to 0.8\textit{a}$_{b}$ for overcontact binaries. Thus, for the circular coplanar planetary orbits with 90\deg  inclination, it can be found that

\begin{equation}
T_{GTV}\simeq\frac{1.8}{2\pi}\frac{a_{b}}{a_{p}}P_{p}.
\end{equation}

Such transit times are usually much longer than the orbital periods of contact binaries. Implications are shown in Fig. 2 as a simulation on the contact binary system W UMa itself, where if we substitute \textit{a}$_{b}$ = 2.45 R$_{\odot}$, \textit{a}$_{p}$ = 260.21 R$_{\odot}$ and \textit{P}$_{p}$ = 1 yr, then we obtain \textit{T}$_{GTV}$ = 0.985 d. It is seen that during the transit of such circumbinary jovian planet, the contact binary inside revolve about three times (0.985 / 0.3336 $\simeq$ 3). In the case of coplanar orbits of the planet and binary, the shape of the transit light curve would be like 3 times $\omega$'s side by side where each $\omega$ representing the transit light curve in one binary orbital period. In Fig. 2, the light variation of the transit (2a) superimposed on the contact binary light curve (2b) as shown in (2c) would be really difficult to disentangle especially in the presence of magnetic activity disturbances (see Fig. 3a). This may be the reason that no transit observations of any circumbinary planet around contact binaries were claimed so far. However, in the light of Fig. 2c, some small changes in the ground-based photometric observations of contact binaries may well be related with transiting circumbinary planets (see e.g. Fig. 3b).

\section{Detection of LITE Effects of The Circumbinary Planets Around Contact Binaries}

It is well known that the cyclic variations in the \textit{O-C} diagrams formed by ET are the signatures of  light time effect (LITE) of unseen objects around binary systems \citep[see e.g.][]{d07}. The ET precision ($\Delta$\textit{t}) depends not only the precision of a single observation but also on the shape, duration and depth of the eclipse minima \citep[see e.g.][]{d06}. The detectable timing amplitude of a cyclic variation ($\Delta$\textit{T}), on the other hand,  is a function of the mass,  orbital period and orbital inclination of the circumbinary third body. The ($\Delta$\textit{T}) should be at least as good as ($\Delta$\textit{t}).  It was estimated that  $\Delta$\textit{T} $\gtrsim$ $\Delta$\textit{t} is usually between 10 - 20 seconds for contact binaries. The orbital periods \textit{P}$_{3}$ of the circumbinary third bodies (with masses of 1 and 10 Jupiter masses) were estimated for the total systemic mass of 1 and 2 Solar mass, and for $\Delta$\textit{T} = 4, 10, 20, and 40 seconds.  Resulting \textit{P}$_{3}$ values (in year) were listed in Table 1.

\begin{table}[!ht]
\caption{The estimated orbital periods (\textit{P}$_{3}$ in years) of the third body with one and 10 Jupiter masses, in the case of LITE amplitude of 4, 10, 20, and 40 seconds. All estimates are for the binary total mass of one solar mass (\textit{m}$_{1}$ + \textit{m}$_{2}$ = 1 M$_{\odot}$). For \textit{m}$_{1}$ + \textit{m}$_{2}$ = 2 M$_{\odot}$, the orbital periods \textit{P}$_{3}$ of the third body doubles. }
\smallskip
\begin{center}
{\small
\begin{tabular}{ccccc}  
\tableline
\noalign{\smallskip}
M$_{j}$ & $\Delta$\textit{T} = 4 sec & $\Delta$\textit{T} = 10 sec& $\Delta$\textit{T} = 20 sec& $\Delta$\textit{T} = 40 sec\\
\noalign{\smallskip}
\tableline
\noalign{\smallskip}
10 & 0.45& 1&3&9 \\
1&14.4&34&97&273 \\

\noalign{\smallskip}
\tableline\
\end{tabular}
}
\end{center}
\end{table}


\begin{figure}
\begin{center}

{\includegraphics[bb= 75 360 344 712, width=5 cm]{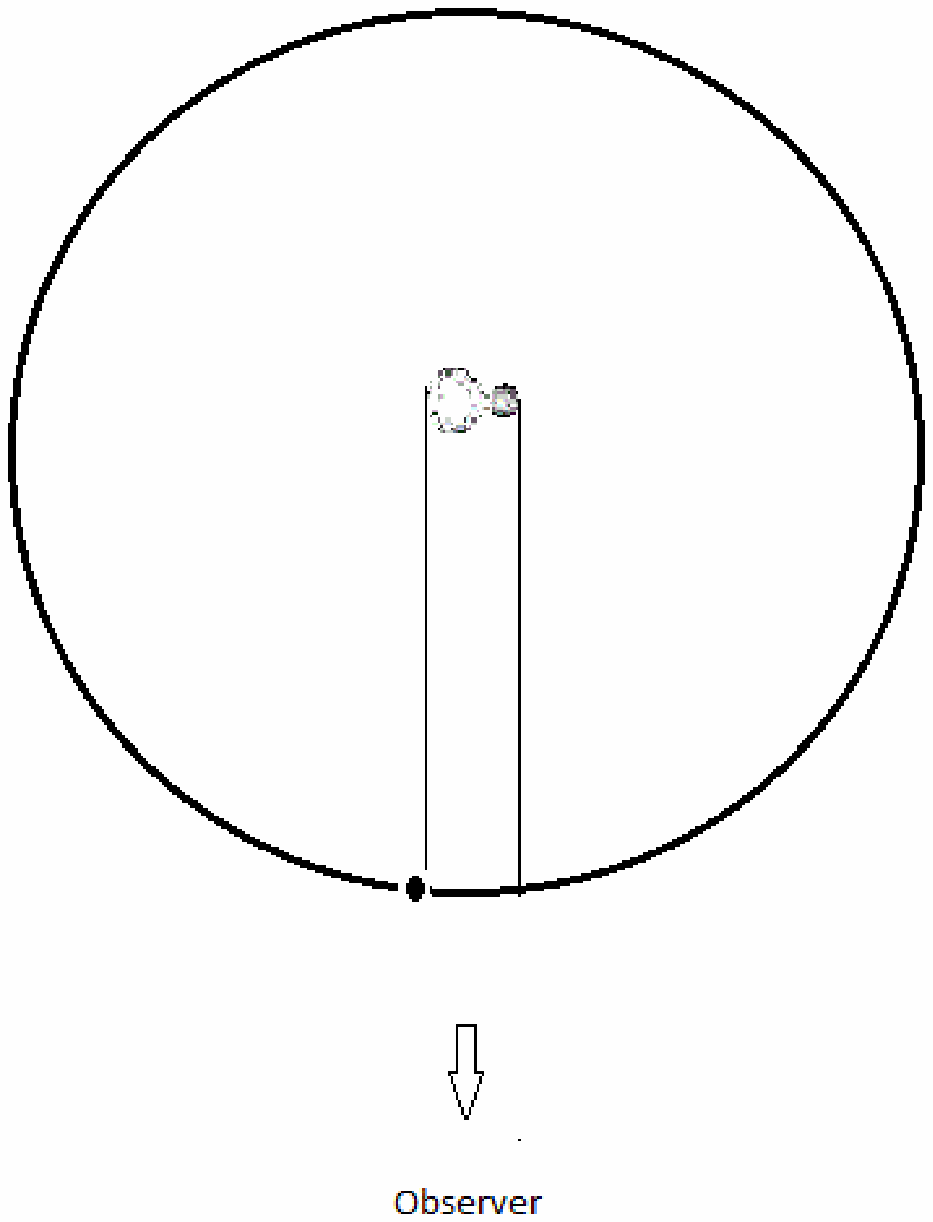}}

\end{center}
\caption{Transit geometry of a circumbinary planet around a contact binary. The planet orbit is expected coplanar with the binary orbit.} \label{fig2}
\end{figure}



\begin{figure}
\begin{center}

{\includegraphics[bb= 12 436 412 766, width=10 cm]{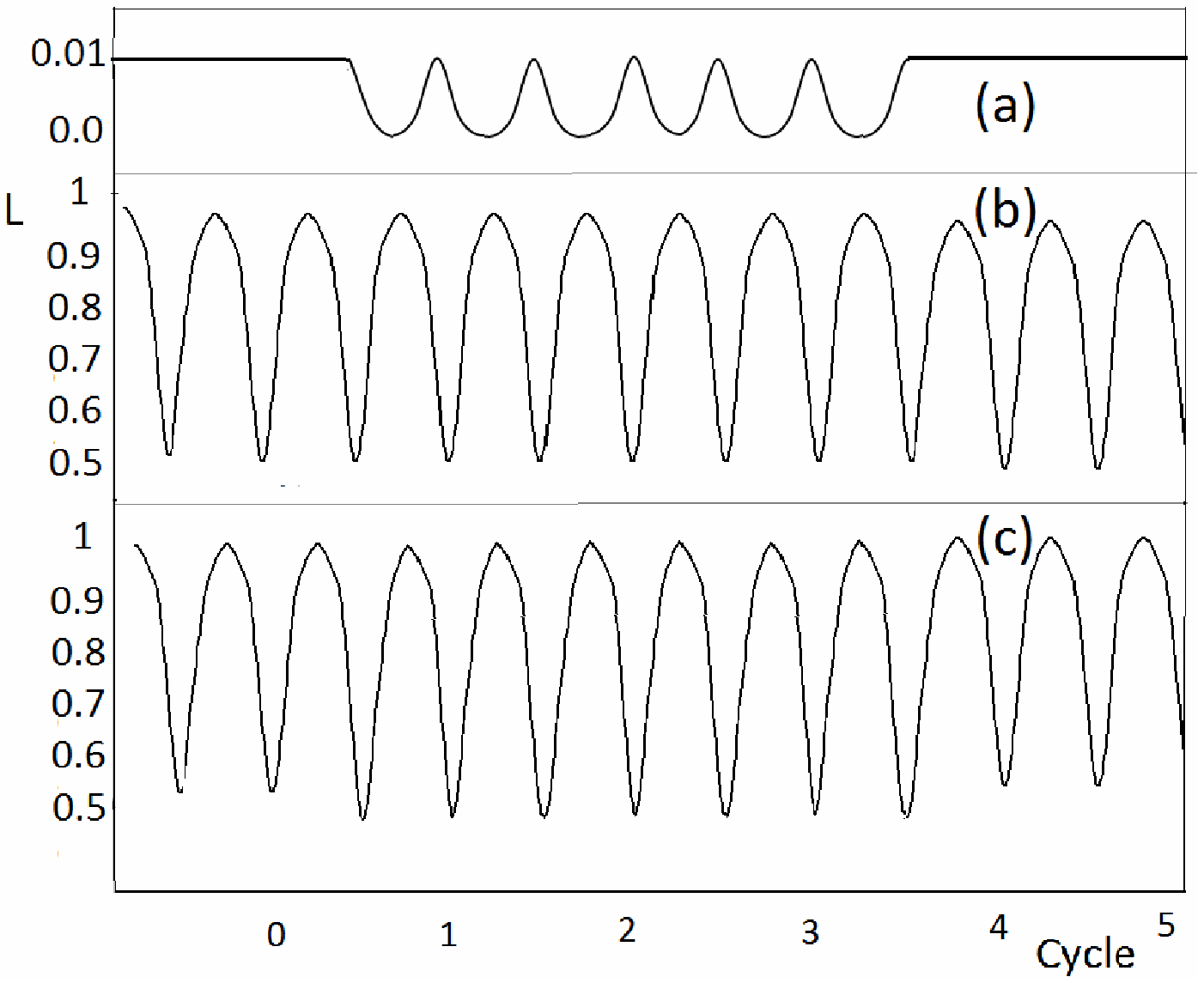}}

\end{center}
\caption{(a) A transit light curve of a circumbinary planet around W UMa system, (b) continuous light variation of W UMa for many cycles, and (c) the combined light variation of W UMa system with transiting circumbinary jovian planet.} \label{fig2}
\end{figure}


\section{Results and Discussion}

The light changes of contact binaries with a possible transiting circumbinary jovian planet was simulated. It is concluded that if the transiting circumbinary planets are present around contact binaries, the relatively large intrinsic light variability, long exposure times and the complexity of the combined low amplitude light variation formed by the transits of the circumbinary planets do not allow yet the disentangling of the transit light curves. 

On the other hand, the ET precision decreases by the intrinsic variability caused by the enhanced magnetic activity of late type G, K, and M- type binary components, and thus the scatter in the \textit{O-C} diagrams increases. The major scatter in the light curves and the \textit{O-C} diagrams are caused by the starspot activity. The time shift in ET (which can be as large as 30 minutes) caused by starspots is proportional to the photometric wave amplitude, while inversely proportional to the depth of the eclipse minima. The starspot effect in the \textit{O-C} diagram can be minimised by using the average of the minimum I and minimum II times, or before forming the \textit{O-C} diagram, the time shifts in ET's can be corrected by spot modelling.
 
Due to relatively large scatter in the \textit{O-C} diagrams of contact binaries, the detectable LITE amplitudes ($\Delta$\textit{T}) would be relatively large and thus, as Table 1 implies, in practice, only the very high mass planets about $m$ = 10 M$_{j}$ are detectable around contact binaries by using ET.


\begin{figure}
\begin{center}

{\includegraphics[bb= 69 542 525 767, width=13 cm]{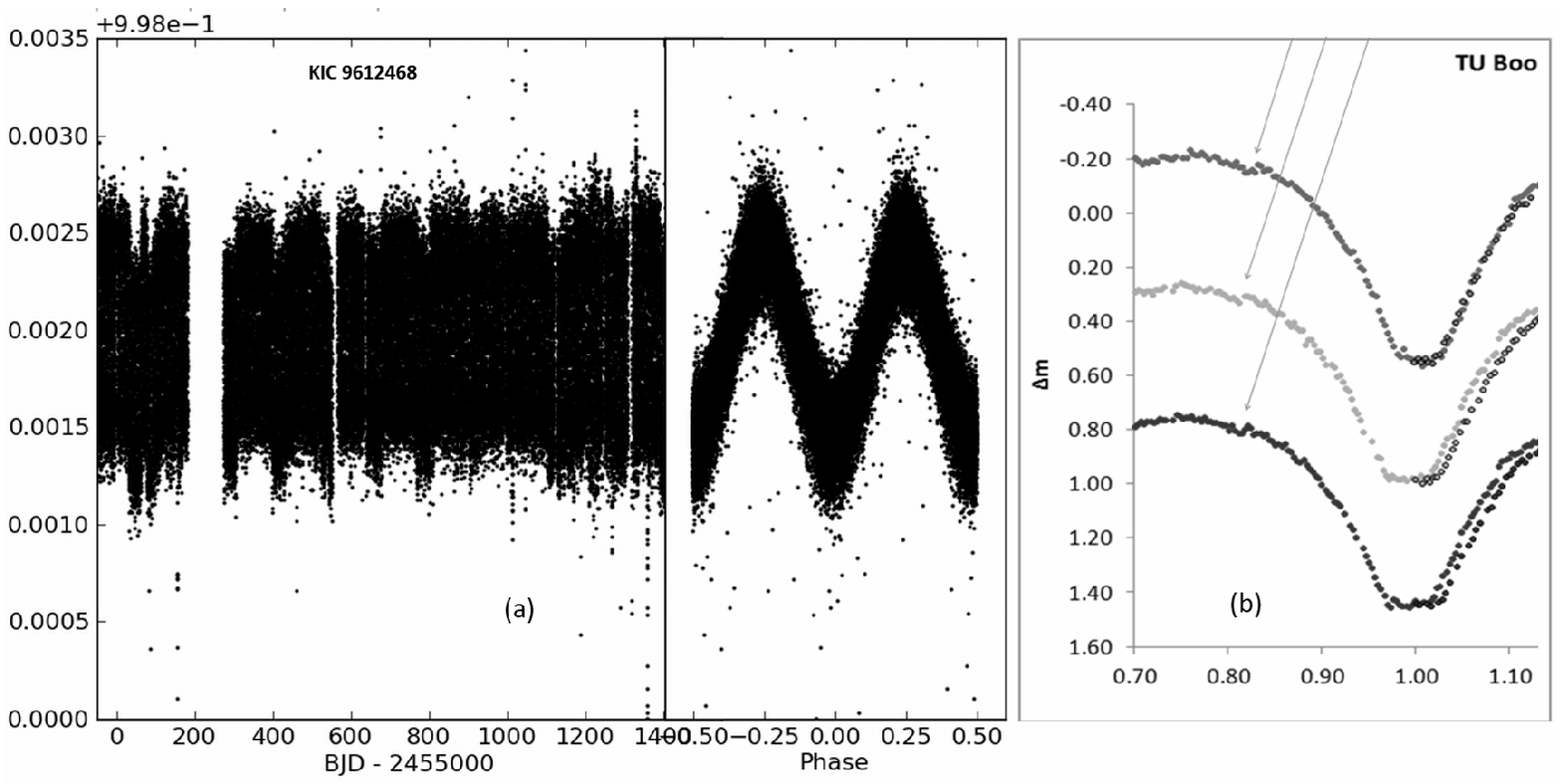}}

\end{center}
\caption{(a) Magnetic activity effect as the variation of the light levels on the light curves of KIC 9612468 extracted from \citet {d09}. (b) Probable transit egress of a circumbinary planet on the light curves of contact binary TU Boo (extracted from \citet {d08}).} \label{fig2}
\end{figure}



\begin{thebibliography}{}

\bibitem[\protect\citeauthoryear{Armstrong et al.}{2013}]{d10}Armstrong, D., Martin, D. V., Brown, G., et al. 2013, $MNRAS$, 434, 3047

\bibitem[\protect\citeauthoryear{Demircan}{1997}]{d07}Demircan, O. 1997, Communications of the Konkoly Observatory, No. 100 (Vol. XII, 2), p. 397-406
 
\bibitem[\protect\citeauthoryear{Dumusque et al.}{2012}]{d01}Dumusque, X., Pepe, F., Lovis, C., et al. 2012, $Nature$, 491, 207

\bibitem[\protect\citeauthoryear{Kley $\&$ Haghighipour}{2014}]{d04}Kley, W. $\&$ Haghighipour, N. 2014, arXiv: 1401.7648 

\bibitem[\protect\citeauthoryear{Kutlualp et al.}{2012}]{d08} Kutlualp, M., Erkan, N., \c{C}i\c{c}ek, C. $\&$ Bulut, A. 2012, 18th Turkish National Astronomy Congress, Malatya, Turkey, Aug 27-31, 2012, 567

\bibitem[\protect\citeauthoryear{Martin $\&$ Triaud}{2014}]{d03}Martin, D.V. $\&$ Triaud, A.H.M.J. 2014, $A$\&$A$, 570, 91

\bibitem[\protect\citeauthoryear{Pr\u{s}a et al.}{2011}]{d09} Pr\u{s}a, A., Batalha, N., Slawson, R. W., et al. 2011, $AJ$, 141, 83, http://keplerebs.villanova.edu

\bibitem[\protect\citeauthoryear{Sybilski et al.}{2010}]{d06} Sybilski, P., Konacki, M. $\&$ Kozlowski, S. 2010, $MNRAS$, 405, 657

\bibitem[\protect\citeauthoryear{Winn}{2010}]{d05} Winn, J. 2010, Exoplanets. The University of Arizona Press, pp 57, 58

\bibitem[\protect\citeauthoryear{Zorotovic $\&$ Schreiber}{2013}]{d02}Zorotovic, M. $\&$ Schreiber, M. R. 2013, $A$\&$A$, 549, 95 




 



\end{thebibliography}
\end{document}